\documentclass[%
 reprint,
 amsmath,amssymb,
 aps,
 prl,
 longbibliography,
]{revtex4-1}
\usepackage{braket}
\usepackage{graphicx}
\usepackage{dcolumn}
\usepackage{bm}
\usepackage{todonotes}
\presetkeys{todonotes}{inline}{}

\begin{document}

\preprint{APS/123-QED}

\title{Quantum-Limited Atomic Receiver in the Electrically Small Regime} 
\author{Kevin C. Cox$^1$}

\author{David H. Meyer$^{1,2}$}%
 
\author{Fredrik K. Fatemi$^1$} 
 
\author{Paul D. Kunz$^1$}

\affiliation{
$^1$U.S. Army Research Laboratory. Adelphi, MD  20783\\
$^2$Dept.~of Physics, University of Maryland. College Park, MD  20742
}%

\date{\today}

\begin{abstract}
 We use a quantum sensor based on thermal Rydberg atoms to receive data encoded in electromagnetic fields in the extreme electrically small regime, with a sensing volume over $10^7$ times smaller than the cube of the electric field wavelength.  We introduce the standard quantum limit for data capacity, and experimentally observe quantum-limited data reception for bandwidths from 10~kHz up to 30~MHz.  In doing this, we provide a useful alternative to classical communication antennas, which become increasingly ineffective when the size of the antenna is significantly smaller than the wavelength of the electromagnetic field. 
\end{abstract}

\maketitle

\newcommand{\BWchu}{BW_\text{Chu}}
\newcommand{\Csql}{C_\text{SQL}} 
\newcommand{\Copt}{C_\text{Opt}} 
\newcommand{\CsqlOpt}{C^*_\text{SQL}}
\newcommand{\vOne}{V_1(f_d)}
\newcommand{\vTwo}{V_1(f_d)}
\newcommand{\vDiff}{V_-(f_d)}
\newcommand{\vNoise}{\Delta V (f_d)}
\newcommand{\Neff}{N_\text{eff}}
\newcommand{\SNR}{\text{SNR}}
\newcommand{\Qchu}{Q_\text{Chu}}

Antennas do not obey Moore's law.  As cutting-edge devices become smaller and smaller, the communication transmitter and receiver antennas present significant size constraints \cite{stutzman_antenna_2012,volakis_antenna_2007}.
This is because fundamental principles limit the performance of a traditional antenna that is significantly smaller than the wavelength of the electromagnetic (EM) field being detected, $\lambda$.  Specifically, a lossless, electrically-small antenna of characteristic radius $a$ is guaranteed to have a quality factor $Q$ greater than the Chu limit,  $\Qchu = \lambda^3/(2 \pi a)^3$, that limits the operation bandwidth to $\BWchu \lesssim f_0/\Qchu$ for carrier frequency $f_0$.  

This limit, pioneered by Chu, Wheeler, Harrington, Mclean and others, influences the design of a wide range of communication technologies using carrier frequencies ranging from DC up to GHz frequencies \cite{volakis_antenna_2007, chu_physical_1948,wheeler_fundamental_1947,harrington_effect_1960,mclean_re-examination_1996}.  Discovery of modest optimizations within the Chu limit constraint is still an active area of research \cite{pfeiffer_fundamental_2017,best_electrically_2015}, as well as exploration into alternative communications technologies, (e.g. based on acoustics or active circuits) that are not subject to the Chu limit \cite{nan_acoustically_2017,sussman-fort_non-foster_2009}.  Inefficient and/or non-impedance-matched designs can relax the Chu limit bandwidth constraint as well.  Here we introduce another alternative path: using a quantum sensor operating at the standard quantum limit to receive classical communications.

In this Letter, we first introduce the quantum-limited data capacity $\Csql$ (a function of signal-to-noise and bandwidth) of a receiver based on a quantum sensor and perform a basic comparison with the Chu limit.  For our experimental parameters, the quantum system nominally gives improvements of over 4-orders of magnitude over the Chu limit.  Second, we experimentally demonstrate reception of signals at the standard quantum limit (SQL) using thermal Rydberg atoms.  By achieving photon-shot-noise-limited readout and increasing the operation bandwidth to be faster than the decoherence rate, we observe a transition from the steady-state electromagnetically-induced transparency (EIT) regime to quantum-limited scaling corresponding to operation at the standard quantum limit for 60 effective atoms. 

A number of recent experiments have used thermal Rydberg atoms for state-of-the-art sensors of electric fields \cite{miller_radio-frequency-modulated_2016,mohapatra_giant_2008, kumar_rydberg-atom_2017, holloway_atom-based_2017, wade_real-time_2017, sedlacek_microwave_2012, anderson_hybrid_2018, anderson_high-resolution_2018}.   Most of these experiments were primarily focused on sensitivity, and operated at bandwidths lower than the decoherence rates set by EIT power broadening, transit broadening, and Doppler broadening, precluding operation at the SQL.  We highlight that for communication purposes, on the other hand, high bandwidth is often the goal, and the fundamental quantum limit can be reached.  Our group and two others have recently demonstrated radio-frequency (rf) communications using a Rydberg atomic sensor \cite{meyer_digital_2018,deb_radio-over-fiber_2018,jiao_atom-based_2018}, but did not reach the SQL, and were not strongly in the electrically small regime, where significant advantages over traditional receiver antennas are apparent.

Atomic sensors are not antennas, at least in the traditional sense.  Traditional antennas are passive devices designed to efficiently convert free space EM waves into signals on a transmission line. On the other hand, atomic electric field sensors often do not absorb net energy from the field, but rather use the atoms and additional laser beams to perform nondestructive sensing.  As an extreme example, consider recent work performing quantum-non-demolition measurements of single microwave photons to stabilize non-classical photon states in a microwave cavity \cite{sayrin_real-time_2011}.  This sensing regime breaks a key assumption behind the Chu limit--namely that of passive, destructive sensing--allowing a quantum sensor using a single atom to operate at an arbitrarily high bandwidth.  

Consider the case where the goal of classical communication is to detect a high rate of data, given by data capacity $C$ measured in bits per second, encoded here in the amplitude modulation of an electric field with a carrier frequency $f_0$.  The achievable $C$ is given by the Shannon-Hartley theorem \cite{shannon_communication_1949}, $C = f_d\times \text{log}_2(1+\SNR^2)$ where $f_d$ is the rate that data symbols are sent, and $\SNR$ is the signal-to-noise ratio, in standard deviation, for detecting the electric field in a measurement window of length $t_d = 1/f_d$.  

\begin{figure}
\begin{centering}
\includegraphics[width=1\linewidth]{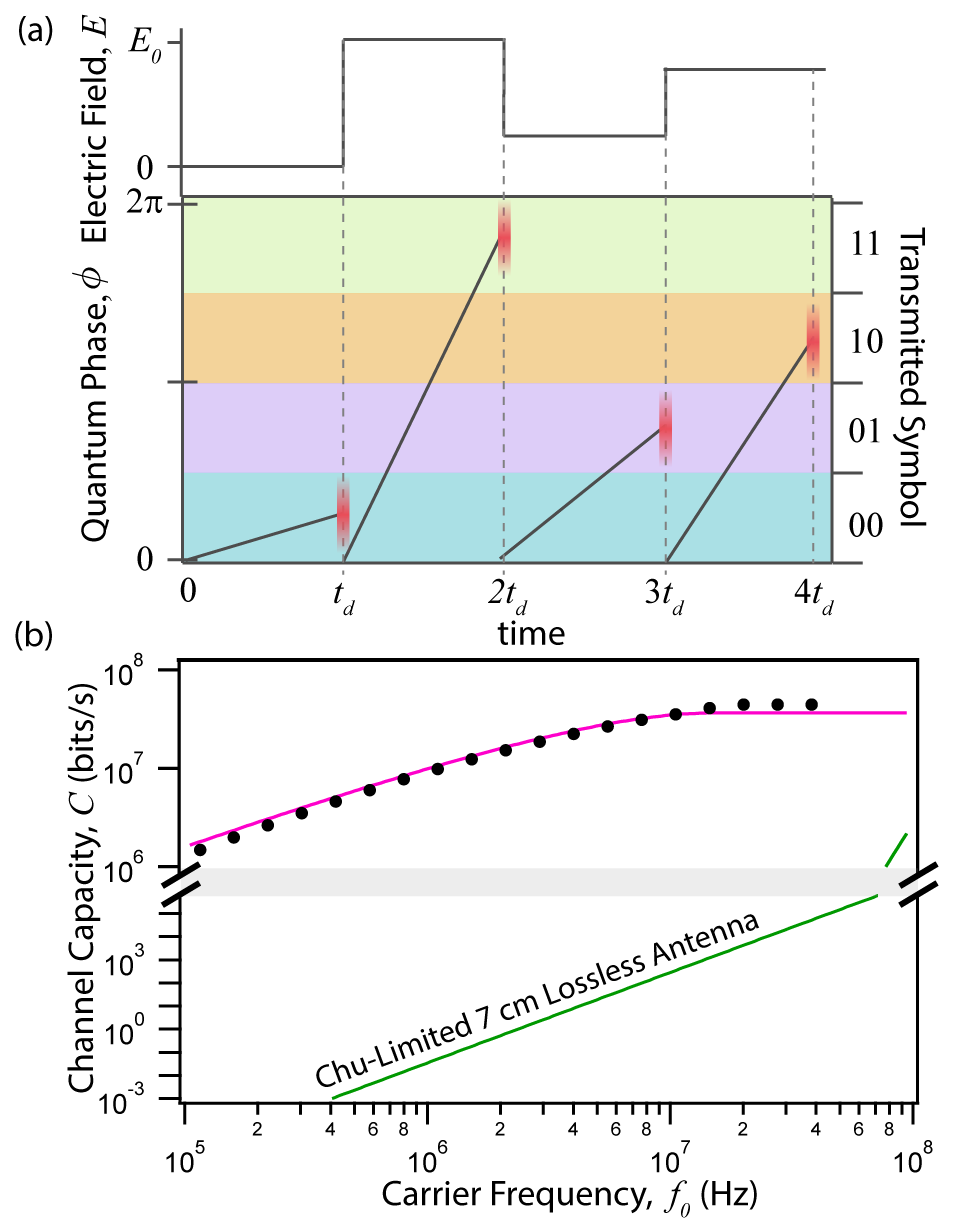}
\par\end{centering}
\caption{\label{fig:Cap}The quantum limit for data capacity. (a)  Data can be sent by encoding information in the strength of the electric field.  The quantum sensor detects symbols by measuring the evolved phase $\phi$ at the end of each period, therefore inferring the transmitted symbol (right axis).  The maximum data capacity is $\Csql = f_d\times \text{log}_2(1+\SNR^2)$~bits/s, where $\SNR$ is the signal-to-noise ratio in standard deviation.  (b)  The Chu limit to channel capacity for an efficient 7~cm classical antenna is shown in green (see text for details). The standard quantum limit $\Csql$ for our experimental parameters is shown in pink.  The directly measured channel capacity of our sensor is shown as black points, where the data rate $f_d$ is restricted to be less than or equal to the carrier frequency $f_0$.}
\label{fig1}

\end{figure}

Any quantum sensor based on 2-level systems observes the applied EM field as an evolution of a quantum phase $\phi$ characterizing a superposition state $\ket{\psi} =\frac{1}{\sqrt{2}} (\ket{g}+e^{i \phi}\ket{e})$ with quantum states $\ket{g}$ and $\ket{e}$.  In the case of low frequency sensing, exclusively studied here, an applied electric field $E$ changes the potential energy $V$ of an atomic dipole, with polarizability $\alpha$, by an amount $V = -\frac{1}{2} \alpha E^2$. The energy shift is observed as a frequency shift $\delta \omega$ of the atomic transition $\delta \omega = V/\hbar$.  In a sensing time $t_d$, $\delta \omega$ accumulates into the evolved quantum phase $\phi = \delta \omega \times t_d$.  When operating with $N$ independent atoms, the individual collapse of atomic wave-functions into either $\ket{e}$ or $\ket{g}$ limits the resolution (given by standard deviation, denoted by $\Delta$) of a measurement of $\phi$ to the standard quantum limit (SQL), $\Delta \phi_{SQL} = 1/\sqrt{N}$. 

Figure 1(a) shows how the SQL limits communication.  For communication, the continuous observable $\phi$ can be broken into a number of discrete binary symbols (e.g. 2 bits, 00 to 11, delineated by color in the figure).  The symbols may be transmitted, as is done here, by changing the amplitude of a static or oscillatory electric field with nominal amplitude $E_0$.  Symbols are received at bandwidth $f_d$ by allowing $\phi$ to linearly evolve into a specific binary state in time $t_d$. In the optimum case, readout is much faster than $t_d$, and quantum noise is observed as an instantaneous uncertainty to each readout of $\phi$ (shown as red uncertainty distributions in Fig. \ref{fig1}(a)).   

Combining the Shannon-Hartley theorem and the SQL, we derive the quantum-limited channel capacity, for $N$ independent atoms, to be
\begin{equation}
\Csql =  f_d\times \text{log}_2\left(1+\frac{\delta \omega^2 \cdot N}{f_d^2}\right).
\end{equation}
$\Csql$ increases with $f_d$ until the argument inside the logarithm becomes approximately 4.92.  This occurs at an optimal (denoted by a star) data transmission rate $f^*_d = 0.505 \times \delta \omega \sqrt{N}$. The corresponding optimal quantum-limited channel capacity is, $\CsqlOpt = 1.16\times \delta \omega \sqrt{N}$.  
To achieve a larger SNR or data capacity, one must increase the atom number or probe an atomic state with a larger polarizability or dipole moment for the carrier frequency used.  

In Fig. 1(b), we present a basic comparison between $\Csql$ and the classical data capacity bound arising from the Chu limit.  To determine the Chu-limited channel capacity, one needs to know both the bandwidth and the SNR of the classical antenna.  Here we consider an efficient classical antenna with maximum Chu-limited data rate $f_d \sim BW_\text{chu}$, whose enclosing sphere \cite{chu_physical_1948} has the same radius, $a$, as that required for our Rb vapor cell ($a= 3.75$~cm).  We consider the classical antenna to be subjected to 50~$\Omega$ Johnson noise at room temperature, and plot for our experimental electric field, 0.8~V/cm. 

The quantum-limited channel capacity $\Csql$ is plotted as a pink line in Fig. \ref{fig1}(b).  The data rate is enforced to be equal to the carrier frequency $f_0$ at low frequencies, but at $f_0=10^7$~Hz, we reach the optimum data rate $f_d = f_0 = f^*_d$. Here the quantum-limited capacity $\Csql$ also reaches the optimum, that is $\CsqlOpt$, and becomes flat.  In the figure, $\Csql$ is plotted for the experimentally observed effective atom number $N_\text{eff} = 63(7)$ and effective observed signal size $\delta \omega_\text{eff} = 2 \pi\times 680(60)$~kHz for the 0.8~V/cm field.  The experimentally measured channel capacity is shown in black points.

The quantum sensor outperforms the efficient electrically-small antenna by a factor of more than $10^4$ at 10~MHz, and the advantage is even more extreme at lower frequencies.  To be clear, there are other methods that surpass the Chu limit such as using inefficient designs \footnote{Traditional AM radio receiver antennas in cars (meter scale) are a good example, which operate with approximately $5\%$ efficiency to reach 10~kHz bandwidth with a $\approx$1~MHz carrier.}, active Non-Foster circuit elements \cite{zhu_broad-bandwidth_2012,sussman-fort_non-foster_2009}, or non-impedance-matched antennas (viable when the field wavelength is long and reflections can be tolerated).  It is also important to note that for many communications applications, external noise sources--blackbody, cosmic, man-made, and atmospheric noise--can dominate the internal receiver noise, be it quantum or classical.  Despite these details that require further exploration, we expect quantum sensors can provide significant benefits in sensitivity and bandwidth for certain applications.

\begin{figure}[tb]
\begin{centering}
\includegraphics[width=1\linewidth]{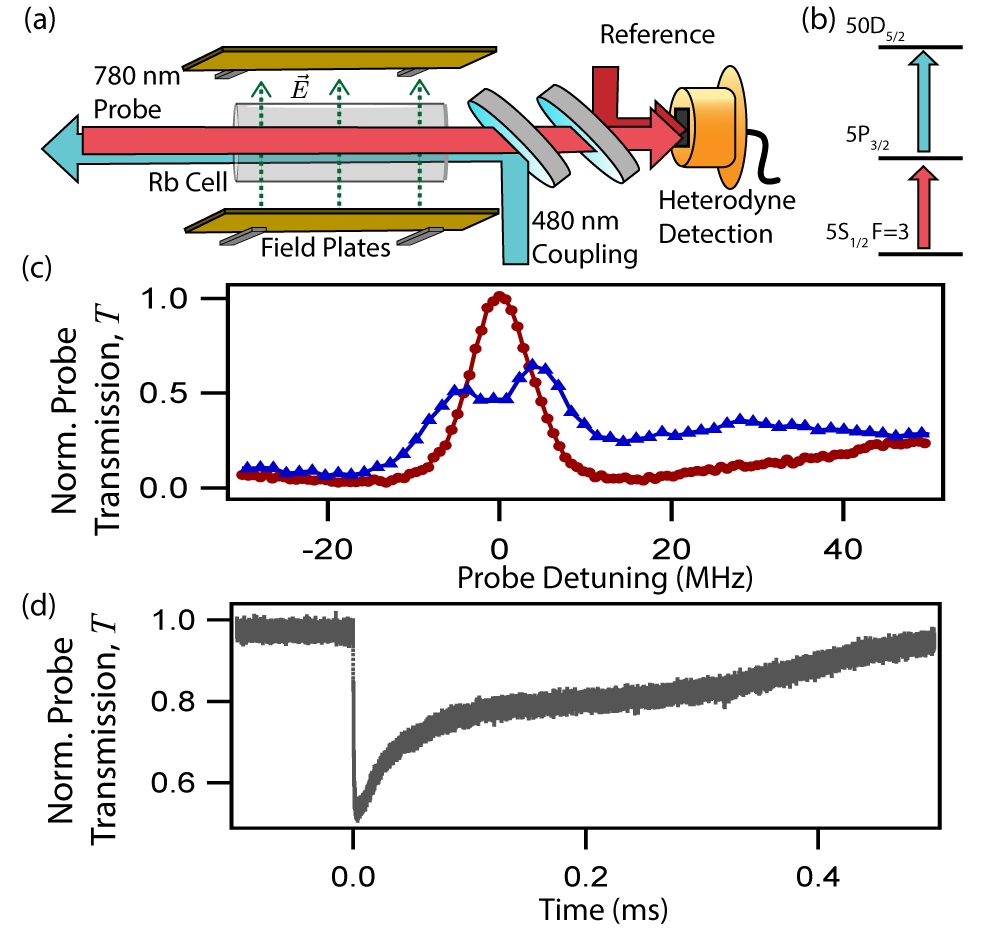}
\par\end{centering}
\caption{\label{fig:timetraces} (a) Experimental setup and (b) Level diagram. (c) When no E-field is applied, we observe a single EIT transmission window (red circles).  Low frequency electric fields cause scalar and tensor Stark shifts, that split the resonance into three peaks (blue triangles).  (d)  At $t=0$, we apply a square pulse to the electric field.  The probe transmission rapidly follows the applied field, but then slowly relaxes over 0.5~ms due to free charges in the glass cell that shield the electric field. }
\label{fig2}
\end{figure}
  
A simplified version of our experimental setup and level diagram is shown in Fig. 2(a) and (b).  Using two parallel plates separated by 60~mm, we apply a transverse low-frequency electric field.  The 480~nm (blue) and 780~nm (red) beams counter-propagate to establish nearly Doppler-free EIT.  In the regime of low-frequency E-fields, we detect the Stark shift of the Rydberg state, $\delta \omega =\frac{1}{2} \alpha  E^2$, where $\alpha$ is the scalar polarizability of the atomic transition, approximately constant for the low frequencies applied here.  The Rydberg state $\ket{R} = \ket{50D_{5/2}}$ splits into three sublevels due to the tensor component of the polarizability: $|m_j|$ = 5/2, $|m_j|$=3/2, $|m_j|$=1/2.  We calculate the polarizabilities of these states to be $\alpha =2\pi\times$ (-42, 42, 212)~MHz/(V/cm)$^2$ respectively.  In Fig. \ref{fig2}(c), we show a plot of the EIT transmission profile with no electric field (red circles), and with an electric field of 0.4~V/cm applied (blue triangles).  The peak splits into the three $m_j$ sublevels.  To detect electric fields, we observe the change in transmission of the 780~nm probe laser through the cell as the EIT resonance frequencies are changed due to the applied electric field.  For high SNR readout, we overlap the 780~nm probe beam with a strong heterodyne local oscillator (LO) detuned by 78.5~MHz.  In contrast to rf systems, this optical heterodyne detection scheme allows readout with zero thermal noise; here we observe photon shot noise to be 6~dB greater than detector noise.  More details about the experimental setup can be found in our previous work, where we demonstrate a digital communication protocol in the microwave domain, with $f_0 =17$~GHz \cite{meyer_digital_2018}.

An example trace, observing the on-resonance probe transmission as the electric field abruptly changes, is shown in Fig. 2(d).  At $t=0$~s, we turn on the electric field from zero (on EIT resonance) to 0.8~V/cm.  The probe transmission (gray line), normalized to the total transmission associated with EIT, drops accordingly, corresponding to the Rydberg state shifting off of 2-photon resonance.  Over 0.5~ms, however, the transmitted signal relaxes to the original value, indicating a relaxation to zero of the electric field observed by the atoms.  This effect has been studied in other vapor-cell based systems and can be attributed to free charges in the glass cell shielding the electric field \cite{miller_radio-frequency-modulated_2016,bason_enhanced_2010}. However, for reasonably high bandwidths of 100~kHz or more, we observe this relaxation effect to be less significant.

\begin{figure}
\begin{centering}
\includegraphics[width=1\linewidth]{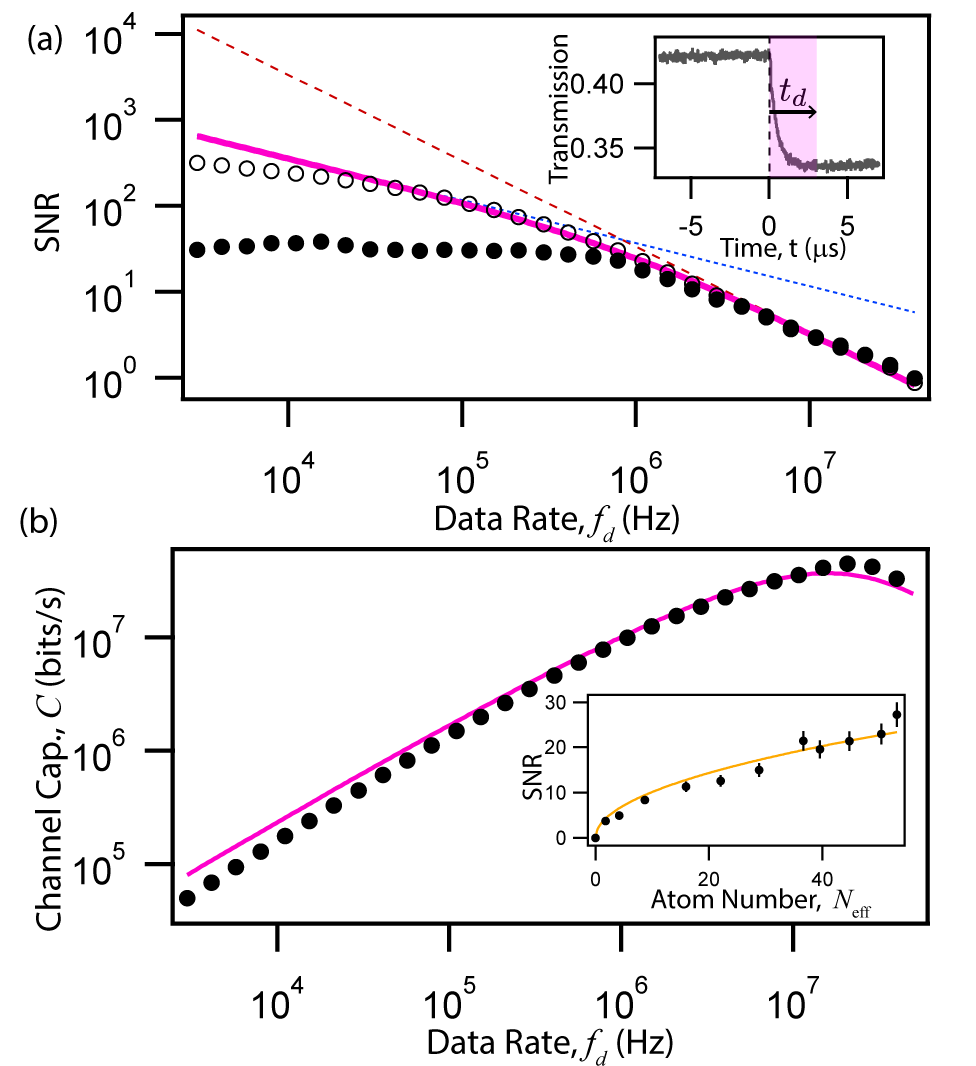}
\par\end{centering}
\caption{\label{fig:SNR}Quantum-limited operation.  (a) We measure the electric field step at t=0 in bandwidth $f_d = 1/t_d$ by averaging the probe transmission through the cell for time $t_d$ (inset). Directly measured SNR for measuring the field vs data bandwidth $f_d$ is shown as black data points.  We subtract out 1/f laser noise to observe the quantum limit over a large range of frequencies (open circles).  At high frequencies, we observe the standard-quantum-limited SNR scaling (red dash).  At lower frequencies, we observe a steady-state, square root scaling of SNR (blue dots).  The data is fit to the complete quantum noise model (pink line).   (b) The SNR data from part a is used to plot channel capacity versus $f_d$.  The quantum limit, for our fitted effective atom number and signal size, is shown as a pink line.  (inset) We plot the SNR for receiving symbols in a bandwidth $f_d = 0.5$~MHz as a function of the effective atom number.  The data is fit to a square-root scaling. }
\label{fig3}
\end{figure}

To calculate the data capacity of our receiver, we measure the SNR for detecting a quantum phase $\phi$ as a function of data rate, $f_d$. Instead of explicitly operating at many different frequencies, we apply a step function in the electric field, as is done in Fig.~2(d) or the inset of Fig. 3(a), and measure the SNR of detecting the step as a function of measurement bandwidth.  Specifically, for each applied step in the field, we average the probe transmission signal in a time window of length $t_d$ (pink window in inset of Fig.~3(a)) placed adjacent to the step.  The electric field is inferred from the outcome of this average.  To change the effective bandwidth, we change the length of the averaging window $t_d = 1/f_d$.  We determine the SNR from the outcome of 100 independent measurements of the electric field.  The resulting SNR for detecting an electric field as a function of data rate/bandwidth is shown as black solid data points in Fig. \ref{fig3}(a).  At frequencies below $10^6$~Hz, $1/f$ laser noise contributes.  We independently measure and subtract off this noise to attain the quantum-limited data shown as open black circles in Fig.~\ref{fig3}(a).

In many quantum sensors, state selective readout means photon shot noise (PSN) is uncorrelated with atomic shot noise \cite{degen_quantum_2017}.  In EIT, on the other hand, scattering of a photon has a one-to-one correspondence with atom wave-function collapse, so that the two are intrinsically linked. Explicitly, the signal-to-noise ratio that we observe is determined by the number of atoms that collapse into the EIT bright state, absorbing and scattering photons out of the probe beam during the communication time $t_d$. This leads to a standard quantum-limited SNR, $\SNR_\text{SQL} = \delta \omega_\text{eff} \cdot t_d \sqrt{Q N}$, where $N$ is the total atom number and $Q$ is the total intrinsic quantum efficiency that includes path losses, technical noise, non-infinite optical depth, as well as the fundamental 50\% efficiency of heterodyne detection.  $Q$ can also be absorbed into an effective atom number $\Neff = Q N$, for which we observe quantum-limited operation.  We also define the effective Stark shift $\delta \omega_\text{eff}$ that accounts for reductions in the signal due to additional decoherence, non-optimal probing, and shielding effects (with associated signal efficiency $Q_\text{sig}$), $\delta \omega_\text{eff} = \delta \omega \times Q_\text{sig}$.

If the symbol period $t_d$ is longer than the coherence time of the dark state in the presence of the electric field, an atom is likely to scatter many times during a single symbol.  In this regime of steady-state (subscript SS) EIT, the SNR is $\SNR_{SS} = \sqrt{N_\text{eff} \cdot t_d/\tau}$, where $\tau$ is the characteristic time for an atom to transition from the dark state to the bright state and scatter a photon.  In Fig.~\ref{fig3}(a), $\SNR_{SS}$ is displayed as a blue dotted line.  If $t_d$ is shorter than $\tau$, atoms collapse, on average, less than once in the symbol period.  In this regime, the signal-to-noise can approach the quantum limit $\SNR_\text{SQL}$.   The SQL for SNR is plotted as a red dashed line in Fig. \ref{fig3}(a). 

We fit our observed SNR in Fig.~3 to a model combining the two SNR limits, $\SNR_\text{SQL}$ and $\SNR_{SS}$. Since the applied Stark shift is larger than the rubidium D2 excited state lifetime ($\Gamma = 2 \pi\times 6$~MHz), we set the scattering rate $1/\tau$ in the model to be the upper bound, $\Gamma/2$ \cite{meyer_digital_2018}.  We allow $\delta \omega_\text{eff}$ and $N_\text{eff}$ to be fit parameters.  The fit is plotted as a pink line in Fig.~3(a).  The fit returns $\delta \omega_\text{eff} = 680(60)$~kHz and $N_\text{eff} = 63(7)$.  From this we deduce $Q_\text{sig} = 0.03$.   Further, the measured optical depth and EIT contrast allows us to approximate the total number of atoms participating in EIT to be of order $10^4$, which gives the total quantum efficiency $Q$ of approximately 0.5\%.  These returned values are in rough agreement with what we predict from known inefficiencies in the current experimental configuration.  We explicitly observe the transition from the steady state PSN regime to the SQL regime at 800~kHz, a frequency governed by $\tau$.  Previous Rydberg electrometry experiments have focused on lower bandwidth sensing, and have not explicitly reached the regime of SQL scaling \cite{kumar_rydberg-atom_2017}.  On the other hand, it is also important to keep in mind that atomic wave-function collapse, resulting in quantum noise in the transmitted light, limits the SNR at all bandwidths, even in the steady-state regime.

In Fig. \ref{fig3}(b) (inset), we  plot the SNR for detecting a symbol in a bandwidth $f_d = 0.5~$MHz as a function of the effective atom number $\Neff$.  Here we adjust $\Neff$ by changing a static electric field, moving the EIT two-photon transition off of resonance. Figure 3(b) shows that the SNR, limited by atomic wavefunction collapse manifesting as PSN, indeed scales as $\sqrt{\Neff}$ (fit displayed as solid orange line).  This scaling is observed in both the steady state and SQL-scaling regimes and can be viewed as either a consequence of random atomic wave-function collapse or photon shot noise.  

In Fig. \ref{fig3}(b), we again plot (black points) the calculated channel capacity $C$, with no noise subtractions, inferred from the measured SNR and data rate of Fig.~3(a) using the Shannon-Hartley theorem.  $\Csql$, using $\delta \omega_\text{eff}$ and $N_\text{eff}$, is shown as a pink line.  As expected $C$ rises approximately linearly until SNR$^2$ = 3.92 around $f_d = 3\times 10^7$~Hz.  At this frequency, we reach the optimum capacity $\CsqlOpt = 4\times 10^7$~bits/s, and the capacity drops at higher data rates.  We highlight that further improvements can only be realized by increasing the effective atom number, the effective polarizability, or adding entanglement between the atoms \cite{degen_quantum_2017,cox_deterministic_2016}.  

More broadly, Fig.~3 associates the performance of our atomic sensor used for classical data reception to the foundational quantum principles governing the system.  This is important, for one, because it sets a fundamental bound--much like the Chu limit for traditional antennas--on the system's capabilities based on the basic resources used.  Second, the ability to relate our receiver's performance to the underlying quantum dynamics also alludes to the potential for Rydberg atomic sensors to extend communication into the quantum regime.  Current work in this area is ongoing \cite{han_coherent_2018,gard_microwave--optical_2017, hattermann_coupling_2017, chen_all-optical_2013,jau_entangling_2016, li_quantum_2016}; we hope that our results further inspire quantum communication tools based on Rydberg vapor cell platforms. 

\section{Acknowledgements}
The authors would like to thank Michael Foss-Feig, Zachary Castillo, David Anderson, Elizabeth Goldschmidt, and Joe Britton for helpful discussions and comments.  This work is supported by the Quantum Science \& Engineering Program of the Office of the Secretary of Defense.

\bibliography{QuantumESA_corr}

\begin{thebibliography}{34}%
\makeatletter
\providecommand \@ifxundefined [1]{%
 \@ifx{#1\undefined}
}%
\providecommand \@ifnum [1]{%
 \ifnum #1\expandafter \@firstoftwo
 \else \expandafter \@secondoftwo
 \fi
}%
\providecommand \@ifx [1]{%
 \ifx #1\expandafter \@firstoftwo
 \else \expandafter \@secondoftwo
 \fi
}%
\providecommand \natexlab [1]{#1}%
\providecommand \enquote  [1]{``#1''}%
\providecommand \bibnamefont  [1]{#1}%
\providecommand \bibfnamefont [1]{#1}%
\providecommand \citenamefont [1]{#1}%
\providecommand \href@noop [0]{\@secondoftwo}%
\providecommand \href [0]{\begingroup \@sanitize@url \@href}%
\providecommand \@href[1]{\@@startlink{#1}\@@href}%
\providecommand \@@href[1]{\endgroup#1\@@endlink}%
\providecommand \@sanitize@url [0]{\catcode `\\12\catcode `\$12\catcode
  `\&12\catcode `\#12\catcode `\^12\catcode `\_12\catcode `\%12\relax}%
\providecommand \@@startlink[1]{}%
\providecommand \@@endlink[0]{}%
\providecommand \url  [0]{\begingroup\@sanitize@url \@url }%
\providecommand \@url [1]{\endgroup\@href {#1}{\urlprefix }}%
\providecommand \urlprefix  [0]{URL }%
\providecommand \Eprint [0]{\href }%
\providecommand \doibase [0]{http://dx.doi.org/}%
\providecommand \selectlanguage [0]{\@gobble}%
\providecommand \bibinfo  [0]{\@secondoftwo}%
\providecommand \bibfield  [0]{\@secondoftwo}%
\providecommand \translation [1]{[#1]}%
\providecommand \BibitemOpen [0]{}%
\providecommand \bibitemStop [0]{}%
\providecommand \bibitemNoStop [0]{.\EOS\space}%
\providecommand \EOS [0]{\spacefactor3000\relax}%
\providecommand \BibitemShut  [1]{\csname bibitem#1\endcsname}%
\let\auto@bib@innerbib\@empty
\bibitem [{\citenamefont {Stutzman}\ and\ \citenamefont
  {Thiele}(2012)}]{stutzman_antenna_2012}%
  \BibitemOpen
  \bibfield  {author} {\bibinfo {author} {\bibfnamefont {Warren~L.}\
  \bibnamefont {Stutzman}}\ and\ \bibinfo {author} {\bibfnamefont {Gary~A.}\
  \bibnamefont {Thiele}},\ }\href@noop {} {\emph {\bibinfo {title} {Antenna
  {Theory} and {Design}}}},\ \bibinfo {edition} {3rd}\ ed.\ (\bibinfo
  {publisher} {Wiley},\ \bibinfo {address} {Hoboken, NJ},\ \bibinfo {year}
  {2012})\BibitemShut {NoStop}%
\bibitem [{\citenamefont {Volakis}(2007)}]{volakis_antenna_2007}%
  \BibitemOpen
  \bibfield  {author} {\bibinfo {author} {\bibfnamefont {John~L.}\ \bibnamefont
  {Volakis}},\ }\href@noop {} {\emph {\bibinfo {title} {Antenna {Engineering}
  {Handbook}}}},\ \bibinfo {edition} {4th}\ ed.\ (\bibinfo  {publisher}
  {McGraw-Hill Education},\ \bibinfo {address} {New York},\ \bibinfo {year}
  {2007})\BibitemShut {NoStop}%
\bibitem [{\citenamefont {Chu}(1948)}]{chu_physical_1948}%
  \BibitemOpen
  \bibfield  {author} {\bibinfo {author} {\bibfnamefont {L.~J.}\ \bibnamefont
  {Chu}},\ }\bibfield  {title} {\enquote {\bibinfo {title} {Physical
  {Limitations} of {Omni}-{Directional} {Antennas}},}\ }\href {\doibase
  10.1063/1.1715038} {\bibfield  {journal} {\bibinfo  {journal} {Journal of
  Applied Physics}\ }\textbf {\bibinfo {volume} {19}},\ \bibinfo {pages}
  {1163--1175} (\bibinfo {year} {1948})}\BibitemShut {NoStop}%
\bibitem [{\citenamefont {Wheeler}(1947)}]{wheeler_fundamental_1947}%
  \BibitemOpen
  \bibfield  {author} {\bibinfo {author} {\bibfnamefont {H.~A.}\ \bibnamefont
  {Wheeler}},\ }\bibfield  {title} {\enquote {\bibinfo {title} {Fundamental
  {Limitations} of {Small} {Antennas}},}\ }\href {\doibase
  10.1109/JRPROC.1947.226199} {\bibfield  {journal} {\bibinfo  {journal}
  {Proceedings of the IRE}\ }\textbf {\bibinfo {volume} {35}},\ \bibinfo
  {pages} {1479--1484} (\bibinfo {year} {1947})}\BibitemShut {NoStop}%
\bibitem [{\citenamefont {Harrington}(1960)}]{harrington_effect_1960}%
  \BibitemOpen
  \bibfield  {author} {\bibinfo {author} {\bibfnamefont {Roger~F.}\
  \bibnamefont {Harrington}},\ }\bibfield  {title} {\enquote {\bibinfo {title}
  {Effect of antenna size on gain, bandwidth, and efficiency},}\ }\href@noop {}
  {\bibfield  {journal} {\bibinfo  {journal} {J. Res. Nat. Bur. Stand}\
  }\textbf {\bibinfo {volume} {64}},\ \bibinfo {pages} {1--12} (\bibinfo {year}
  {1960})}\BibitemShut {NoStop}%
\bibitem [{\citenamefont {McLean}(1996)}]{mclean_re-examination_1996}%
  \BibitemOpen
  \bibfield  {author} {\bibinfo {author} {\bibfnamefont {J.~S.}\ \bibnamefont
  {McLean}},\ }\bibfield  {title} {\enquote {\bibinfo {title} {A re-examination
  of the fundamental limits on the radiation {Q} of electrically small
  antennas},}\ }\href {\doibase 10.1109/8.496253} {\bibfield  {journal}
  {\bibinfo  {journal} {IEEE Transactions on Antennas and Propagation}\
  }\textbf {\bibinfo {volume} {44}},\ \bibinfo {pages} {672--} (\bibinfo {year}
  {1996})}\BibitemShut {NoStop}%
\bibitem [{\citenamefont {Pfeiffer}(2017)}]{pfeiffer_fundamental_2017}%
  \BibitemOpen
  \bibfield  {author} {\bibinfo {author} {\bibfnamefont {C.}~\bibnamefont
  {Pfeiffer}},\ }\bibfield  {title} {\enquote {\bibinfo {title} {Fundamental
  {Efficiency} {Limits} for {Small} {Metallic} {Antennas}},}\ }\href {\doibase
  10.1109/TAP.2017.2670532} {\bibfield  {journal} {\bibinfo  {journal} {IEEE
  Transactions on Antennas and Propagation}\ }\textbf {\bibinfo {volume}
  {65}},\ \bibinfo {pages} {1642--1650} (\bibinfo {year} {2017})}\BibitemShut
  {NoStop}%
\bibitem [{\citenamefont {Best}(2015)}]{best_electrically_2015}%
  \BibitemOpen
  \bibfield  {author} {\bibinfo {author} {\bibfnamefont {S.~R.}\ \bibnamefont
  {Best}},\ }\bibfield  {title} {\enquote {\bibinfo {title} {Electrically
  {Small} {Resonant} {Planar} {Antennas}: {Optimizing} the quality factor and
  bandwidth.}}\ }\href {\doibase 10.1109/MAP.2015.2437271} {\bibfield
  {journal} {\bibinfo  {journal} {IEEE Antennas and Propagation Magazine}\
  }\textbf {\bibinfo {volume} {57}},\ \bibinfo {pages} {38--47} (\bibinfo
  {year} {2015})}\BibitemShut {NoStop}%
\bibitem [{\citenamefont {Nan}\ \emph {et~al.}(2017)\citenamefont {Nan},
  \citenamefont {Lin}, \citenamefont {Gao}, \citenamefont {Matyushov},
  \citenamefont {Yu}, \citenamefont {Chen}, \citenamefont {Sun}, \citenamefont
  {Wei}, \citenamefont {Wang}, \citenamefont {Li}, \citenamefont {Wang},
  \citenamefont {Belkessam}, \citenamefont {Guo}, \citenamefont {Chen},
  \citenamefont {Zhou}, \citenamefont {Qian}, \citenamefont {Hui},
  \citenamefont {Rinaldi}, \citenamefont {McConney}, \citenamefont {Howe},
  \citenamefont {Hu}, \citenamefont {Jones}, \citenamefont {Brown},\ and\
  \citenamefont {Sun}}]{nan_acoustically_2017}%
  \BibitemOpen
  \bibfield  {author} {\bibinfo {author} {\bibfnamefont {Tianxiang}\
  \bibnamefont {Nan}}, \bibinfo {author} {\bibfnamefont {Hwaider}\ \bibnamefont
  {Lin}}, \bibinfo {author} {\bibfnamefont {Yuan}\ \bibnamefont {Gao}},
  \bibinfo {author} {\bibfnamefont {Alexei}\ \bibnamefont {Matyushov}},
  \bibinfo {author} {\bibfnamefont {Guoliang}\ \bibnamefont {Yu}}, \bibinfo
  {author} {\bibfnamefont {Huaihao}\ \bibnamefont {Chen}}, \bibinfo {author}
  {\bibfnamefont {Neville}\ \bibnamefont {Sun}}, \bibinfo {author}
  {\bibfnamefont {Shengjun}\ \bibnamefont {Wei}}, \bibinfo {author}
  {\bibfnamefont {Zhiguang}\ \bibnamefont {Wang}}, \bibinfo {author}
  {\bibfnamefont {Menghui}\ \bibnamefont {Li}}, \bibinfo {author}
  {\bibfnamefont {Xinjun}\ \bibnamefont {Wang}}, \bibinfo {author}
  {\bibfnamefont {Amine}\ \bibnamefont {Belkessam}}, \bibinfo {author}
  {\bibfnamefont {Rongdi}\ \bibnamefont {Guo}}, \bibinfo {author}
  {\bibfnamefont {Brian}\ \bibnamefont {Chen}}, \bibinfo {author}
  {\bibfnamefont {James}\ \bibnamefont {Zhou}}, \bibinfo {author}
  {\bibfnamefont {Zhenyun}\ \bibnamefont {Qian}}, \bibinfo {author}
  {\bibfnamefont {Yu}~\bibnamefont {Hui}}, \bibinfo {author} {\bibfnamefont
  {Matteo}\ \bibnamefont {Rinaldi}}, \bibinfo {author} {\bibfnamefont
  {Michael~E.}\ \bibnamefont {McConney}}, \bibinfo {author} {\bibfnamefont
  {Brandon~M.}\ \bibnamefont {Howe}}, \bibinfo {author} {\bibfnamefont
  {Zhongqiang}\ \bibnamefont {Hu}}, \bibinfo {author} {\bibfnamefont {John~G.}\
  \bibnamefont {Jones}}, \bibinfo {author} {\bibfnamefont {Gail~J.}\
  \bibnamefont {Brown}}, \ and\ \bibinfo {author} {\bibfnamefont {Nian~Xiang}\
  \bibnamefont {Sun}},\ }\bibfield  {title} {\enquote {\bibinfo {title}
  {Acoustically actuated ultra-compact {NEMS} magnetoelectric antennas},}\
  }\href {\doibase 10.1038/s41467-017-00343-8} {\bibfield  {journal} {\bibinfo
  {journal} {Nature Communications}\ }\textbf {\bibinfo {volume} {8}},\
  \bibinfo {pages} {296} (\bibinfo {year} {2017})}\BibitemShut {NoStop}%
\bibitem [{\citenamefont {Sussman-Fort}\ and\ \citenamefont
  {Rudish}(2009)}]{sussman-fort_non-foster_2009}%
  \BibitemOpen
  \bibfield  {author} {\bibinfo {author} {\bibfnamefont {S.~E.}\ \bibnamefont
  {Sussman-Fort}}\ and\ \bibinfo {author} {\bibfnamefont {R.~M.}\ \bibnamefont
  {Rudish}},\ }\bibfield  {title} {\enquote {\bibinfo {title} {Non-{Foster}
  {Impedance} {Matching} of {Electrically}-{Small} {Antennas}},}\ }\href
  {\doibase 10.1109/TAP.2009.2024494} {\bibfield  {journal} {\bibinfo
  {journal} {IEEE Transactions on Antennas and Propagation}\ }\textbf {\bibinfo
  {volume} {57}},\ \bibinfo {pages} {2230--2241} (\bibinfo {year}
  {2009})}\BibitemShut {NoStop}%
\bibitem [{\citenamefont {Miller}\ \emph {et~al.}(2016)\citenamefont {Miller},
  \citenamefont {Anderson},\ and\ \citenamefont
  {Raithel}}]{miller_radio-frequency-modulated_2016}%
  \BibitemOpen
  \bibfield  {author} {\bibinfo {author} {\bibfnamefont {S.~A.}\ \bibnamefont
  {Miller}}, \bibinfo {author} {\bibfnamefont {D.~A.}\ \bibnamefont
  {Anderson}}, \ and\ \bibinfo {author} {\bibfnamefont {G.}~\bibnamefont
  {Raithel}},\ }\bibfield  {title} {\enquote {\bibinfo {title}
  {Radio-frequency-modulated {Rydberg} states in a vapor cell},}\ }\href
  {\doibase 10.1088/1367-2630/18/5/053017} {\bibfield  {journal} {\bibinfo
  {journal} {New J. Phys.}\ }\textbf {\bibinfo {volume} {18}},\ \bibinfo
  {pages} {053017} (\bibinfo {year} {2016})}\BibitemShut {NoStop}%
\bibitem [{\citenamefont {Mohapatra}\ \emph {et~al.}(2008)\citenamefont
  {Mohapatra}, \citenamefont {Bason}, \citenamefont {Butscher}, \citenamefont
  {Weatherill},\ and\ \citenamefont {Adams}}]{mohapatra_giant_2008}%
  \BibitemOpen
  \bibfield  {author} {\bibinfo {author} {\bibfnamefont {Ashok~K.}\
  \bibnamefont {Mohapatra}}, \bibinfo {author} {\bibfnamefont {Mark~G.}\
  \bibnamefont {Bason}}, \bibinfo {author} {\bibfnamefont {Bj{\"o}rn}\
  \bibnamefont {Butscher}}, \bibinfo {author} {\bibfnamefont {Kevin~J.}\
  \bibnamefont {Weatherill}}, \ and\ \bibinfo {author} {\bibfnamefont
  {Charles~S.}\ \bibnamefont {Adams}},\ }\bibfield  {title} {\enquote {\bibinfo
  {title} {A giant electro-optic effect using polarizable dark states},}\
  }\href {\doibase 10.1038/nphys1091} {\bibfield  {journal} {\bibinfo
  {journal} {Nature Physics}\ }\textbf {\bibinfo {volume} {4}},\ \bibinfo
  {pages} {890--894} (\bibinfo {year} {2008})}\BibitemShut {NoStop}%
\bibitem [{\citenamefont {Kumar}\ \emph {et~al.}(2017)\citenamefont {Kumar},
  \citenamefont {Fan}, \citenamefont {K{\"u}bler}, \citenamefont {Jahangiri},\
  and\ \citenamefont {Shaffer}}]{kumar_rydberg-atom_2017}%
  \BibitemOpen
  \bibfield  {author} {\bibinfo {author} {\bibfnamefont {Santosh}\ \bibnamefont
  {Kumar}}, \bibinfo {author} {\bibfnamefont {Haoquan}\ \bibnamefont {Fan}},
  \bibinfo {author} {\bibfnamefont {Harald}\ \bibnamefont {K{\"u}bler}},
  \bibinfo {author} {\bibfnamefont {Akbar~J.}\ \bibnamefont {Jahangiri}}, \
  and\ \bibinfo {author} {\bibfnamefont {James~P.}\ \bibnamefont {Shaffer}},\
  }\bibfield  {title} {\enquote {\bibinfo {title} {Rydberg-atom based
  radio-frequency electrometry using frequency modulation spectroscopy in room
  temperature vapor cells},}\ }\href {\doibase 10.1364/OE.25.008625} {\bibfield
   {journal} {\bibinfo  {journal} {Opt. Express, OE}\ }\textbf {\bibinfo
  {volume} {25}},\ \bibinfo {pages} {8625--8637} (\bibinfo {year}
  {2017})}\BibitemShut {NoStop}%
\bibitem [{\citenamefont {Holloway}\ \emph {et~al.}(2017)\citenamefont
  {Holloway}, \citenamefont {Simons}, \citenamefont {Gordon}, \citenamefont
  {Wilson}, \citenamefont {Cooke}, \citenamefont {Anderson},\ and\
  \citenamefont {Raithel}}]{holloway_atom-based_2017}%
  \BibitemOpen
  \bibfield  {author} {\bibinfo {author} {\bibfnamefont {C.~L.}\ \bibnamefont
  {Holloway}}, \bibinfo {author} {\bibfnamefont {M.~T.}\ \bibnamefont
  {Simons}}, \bibinfo {author} {\bibfnamefont {J.~A.}\ \bibnamefont {Gordon}},
  \bibinfo {author} {\bibfnamefont {P.~F.}\ \bibnamefont {Wilson}}, \bibinfo
  {author} {\bibfnamefont {C.~M.}\ \bibnamefont {Cooke}}, \bibinfo {author}
  {\bibfnamefont {D.~A.}\ \bibnamefont {Anderson}}, \ and\ \bibinfo {author}
  {\bibfnamefont {G.}~\bibnamefont {Raithel}},\ }\bibfield  {title} {\enquote
  {\bibinfo {title} {Atom-{Based} {RF} {Electric} {Field} {Metrology}: {From}
  {Self}-{Calibrated} {Measurements} to {Subwavelength} and {Near}-{Field}
  {Imaging}},}\ }\href {\doibase 10.1109/TEMC.2016.2644616} {\bibfield
  {journal} {\bibinfo  {journal} {IEEE Transactions on Electromagnetic
  Compatibility}\ }\textbf {\bibinfo {volume} {59}},\ \bibinfo {pages}
  {717--728} (\bibinfo {year} {2017})}\BibitemShut {NoStop}%
\bibitem [{\citenamefont {Wade}\ \emph {et~al.}(2017)\citenamefont {Wade},
  \citenamefont {{\v S}ibali{\'c}}, \citenamefont {Melo}, \citenamefont
  {Kondo}, \citenamefont {Adams},\ and\ \citenamefont
  {Weatherill}}]{wade_real-time_2017}%
  \BibitemOpen
  \bibfield  {author} {\bibinfo {author} {\bibfnamefont {C.~G.}\ \bibnamefont
  {Wade}}, \bibinfo {author} {\bibfnamefont {N.}~\bibnamefont {{\v
  S}ibali{\'c}}}, \bibinfo {author} {\bibfnamefont {N.~R.~de}\ \bibnamefont
  {Melo}}, \bibinfo {author} {\bibfnamefont {J.~M.}\ \bibnamefont {Kondo}},
  \bibinfo {author} {\bibfnamefont {C.~S.}\ \bibnamefont {Adams}}, \ and\
  \bibinfo {author} {\bibfnamefont {K.~J.}\ \bibnamefont {Weatherill}},\
  }\bibfield  {title} {\enquote {\bibinfo {title} {Real-time near-field
  terahertz imaging with atomic optical fluorescence},}\ }\href {\doibase
  10.1038/nphoton.2016.214} {\bibfield  {journal} {\bibinfo  {journal} {Nature
  Photonics}\ }\textbf {\bibinfo {volume} {11}},\ \bibinfo {pages} {40--43}
  (\bibinfo {year} {2017})}\BibitemShut {NoStop}%
\bibitem [{\citenamefont {Sedlacek}\ \emph {et~al.}(2012)\citenamefont
  {Sedlacek}, \citenamefont {Schwettmann}, \citenamefont {K{\"u}bler},
  \citenamefont {L{\"o}w}, \citenamefont {Pfau},\ and\ \citenamefont
  {Shaffer}}]{sedlacek_microwave_2012}%
  \BibitemOpen
  \bibfield  {author} {\bibinfo {author} {\bibfnamefont {Jonathon~A.}\
  \bibnamefont {Sedlacek}}, \bibinfo {author} {\bibfnamefont {Arne}\
  \bibnamefont {Schwettmann}}, \bibinfo {author} {\bibfnamefont {Harald}\
  \bibnamefont {K{\"u}bler}}, \bibinfo {author} {\bibfnamefont {Robert}\
  \bibnamefont {L{\"o}w}}, \bibinfo {author} {\bibfnamefont {Tilman}\
  \bibnamefont {Pfau}}, \ and\ \bibinfo {author} {\bibfnamefont {James~P.}\
  \bibnamefont {Shaffer}},\ }\bibfield  {title} {\enquote {\bibinfo {title}
  {Microwave electrometry with {Rydberg} atoms in a vapour cell using bright
  atomic resonances},}\ }\href {\doibase 10.1038/nphys2423} {\bibfield
  {journal} {\bibinfo  {journal} {Nat Phys}\ }\textbf {\bibinfo {volume} {8}},\
  \bibinfo {pages} {819--824} (\bibinfo {year} {2012})}\BibitemShut {NoStop}%
\bibitem [{\citenamefont {Anderson}\ \emph
  {et~al.}(2018{\natexlab{a}})\citenamefont {Anderson}, \citenamefont
  {Paradis},\ and\ \citenamefont {Raithel}}]{anderson_hybrid_2018}%
  \BibitemOpen
  \bibfield  {author} {\bibinfo {author} {\bibfnamefont {David~A.}\
  \bibnamefont {Anderson}}, \bibinfo {author} {\bibfnamefont {Eric~G.}\
  \bibnamefont {Paradis}}, \ and\ \bibinfo {author} {\bibfnamefont {Georg}\
  \bibnamefont {Raithel}},\ }\bibfield  {title} {\enquote {\bibinfo {title} {A
  hybrid polarization-selective atomic sensor for radio-frequency field
  detection with a passive resonant-cavity field amplifier},}\ }\href
  {http://arxiv.org/abs/1805.00412} {\bibfield  {journal} {\bibinfo  {journal}
  {arXiv:1805.00412 [physics, physics:quant-ph]}\ } (\bibinfo {year}
  {2018}{\natexlab{a}})},\ \bibinfo {note} {arXiv: 1805.00412}\BibitemShut
  {NoStop}%
\bibitem [{\citenamefont {Anderson}\ \emph
  {et~al.}(2018{\natexlab{b}})\citenamefont {Anderson}, \citenamefont
  {Paradis}, \citenamefont {Raithel}, \citenamefont {Sapiro},\ and\
  \citenamefont {Holloway}}]{anderson_high-resolution_2018}%
  \BibitemOpen
  \bibfield  {author} {\bibinfo {author} {\bibfnamefont {David~A.}\
  \bibnamefont {Anderson}}, \bibinfo {author} {\bibfnamefont {Eric}\
  \bibnamefont {Paradis}}, \bibinfo {author} {\bibfnamefont {Georg}\
  \bibnamefont {Raithel}}, \bibinfo {author} {\bibfnamefont {Rachel~E.}\
  \bibnamefont {Sapiro}}, \ and\ \bibinfo {author} {\bibfnamefont
  {Christopher~L.}\ \bibnamefont {Holloway}},\ }\bibfield  {title} {\enquote
  {\bibinfo {title} {High-resolution antenna near-field imaging and sub-{THz}
  measurements with a small atomic vapor-cell sensing element},}\ }\href
  {http://arxiv.org/abs/1804.09789} {\bibfield  {journal} {\bibinfo  {journal}
  {arXiv:1804.09789 [physics, physics:quant-ph]}\ } (\bibinfo {year}
  {2018}{\natexlab{b}})},\ \bibinfo {note} {arXiv: 1804.09789}\BibitemShut
  {NoStop}%
\bibitem [{\citenamefont {Meyer}\ \emph {et~al.}(2018)\citenamefont {Meyer},
  \citenamefont {Cox}, \citenamefont {Fatemi},\ and\ \citenamefont
  {Kunz}}]{meyer_digital_2018}%
  \BibitemOpen
  \bibfield  {author} {\bibinfo {author} {\bibfnamefont {David~H.}\
  \bibnamefont {Meyer}}, \bibinfo {author} {\bibfnamefont {Kevin~C.}\
  \bibnamefont {Cox}}, \bibinfo {author} {\bibfnamefont {Fredrik~K.}\
  \bibnamefont {Fatemi}}, \ and\ \bibinfo {author} {\bibfnamefont {Paul~D.}\
  \bibnamefont {Kunz}},\ }\bibfield  {title} {\enquote {\bibinfo {title}
  {Digital communication with {Rydberg} atoms and amplitude-modulated microwave
  fields},}\ }\href {\doibase 10.1063/1.5028357} {\bibfield  {journal}
  {\bibinfo  {journal} {Applied Physics Letters}\ }\textbf {\bibinfo {volume}
  {112}},\ \bibinfo {pages} {211108} (\bibinfo {year} {2018})}\BibitemShut
  {NoStop}%
\bibitem [{\citenamefont {Deb}\ and\ \citenamefont
  {Kj{\ae}rgaard}(2018)}]{deb_radio-over-fiber_2018}%
  \BibitemOpen
  \bibfield  {author} {\bibinfo {author} {\bibfnamefont {A.~B.}\ \bibnamefont
  {Deb}}\ and\ \bibinfo {author} {\bibfnamefont {N.}~\bibnamefont
  {Kj{\ae}rgaard}},\ }\bibfield  {title} {\enquote {\bibinfo {title}
  {Radio-over-fiber using an optical antenna based on {Rydberg} states of
  atoms},}\ }\href {\doibase 10.1063/1.5031033} {\bibfield  {journal} {\bibinfo
   {journal} {Appl. Phys. Lett.}\ }\textbf {\bibinfo {volume} {112}},\ \bibinfo
  {pages} {211106} (\bibinfo {year} {2018})}\BibitemShut {NoStop}%
\bibitem [{\citenamefont {Jiao}\ \emph {et~al.}(2018)\citenamefont {Jiao},
  \citenamefont {Han}, \citenamefont {Fan}, \citenamefont {Raithel},
  \citenamefont {Zhao},\ and\ \citenamefont {Jia}}]{jiao_atom-based_2018}%
  \BibitemOpen
  \bibfield  {author} {\bibinfo {author} {\bibfnamefont {Yuechun}\ \bibnamefont
  {Jiao}}, \bibinfo {author} {\bibfnamefont {Xiaoxuan}\ \bibnamefont {Han}},
  \bibinfo {author} {\bibfnamefont {Jiabei}\ \bibnamefont {Fan}}, \bibinfo
  {author} {\bibfnamefont {Georg}\ \bibnamefont {Raithel}}, \bibinfo {author}
  {\bibfnamefont {Jianming}\ \bibnamefont {Zhao}}, \ and\ \bibinfo {author}
  {\bibfnamefont {Suotang}\ \bibnamefont {Jia}},\ }\bibfield  {title} {\enquote
  {\bibinfo {title} {Atom-based quantum receiver for amplitude- and
  frequency-modulated baseband signals in high-frequency radio
  communication},}\ }\href {http://arxiv.org/abs/1804.07044} {\  (\bibinfo
  {year} {2018})},\ \bibinfo {note} {arXiv: 1804.07044}\BibitemShut {NoStop}%
\bibitem [{\citenamefont {Sayrin}\ \emph {et~al.}(2011)\citenamefont {Sayrin},
  \citenamefont {Dotsenko}, \citenamefont {Zhou}, \citenamefont {Peaudecerf},
  \citenamefont {Rybarczyk}, \citenamefont {Gleyzes}, \citenamefont {Rouchon},
  \citenamefont {Mirrahimi}, \citenamefont {Amini}, \citenamefont {Brune},
  \citenamefont {Raimond},\ and\ \citenamefont
  {Haroche}}]{sayrin_real-time_2011}%
  \BibitemOpen
  \bibfield  {author} {\bibinfo {author} {\bibfnamefont {Cl{\'e}ment}\
  \bibnamefont {Sayrin}}, \bibinfo {author} {\bibfnamefont {Igor}\ \bibnamefont
  {Dotsenko}}, \bibinfo {author} {\bibfnamefont {Xingxing}\ \bibnamefont
  {Zhou}}, \bibinfo {author} {\bibfnamefont {Bruno}\ \bibnamefont
  {Peaudecerf}}, \bibinfo {author} {\bibfnamefont {Th{\'e}o}\ \bibnamefont
  {Rybarczyk}}, \bibinfo {author} {\bibfnamefont {S{\'e}bastien}\ \bibnamefont
  {Gleyzes}}, \bibinfo {author} {\bibfnamefont {Pierre}\ \bibnamefont
  {Rouchon}}, \bibinfo {author} {\bibfnamefont {Mazyar}\ \bibnamefont
  {Mirrahimi}}, \bibinfo {author} {\bibfnamefont {Hadis}\ \bibnamefont
  {Amini}}, \bibinfo {author} {\bibfnamefont {Michel}\ \bibnamefont {Brune}},
  \bibinfo {author} {\bibfnamefont {Jean-Michel}\ \bibnamefont {Raimond}}, \
  and\ \bibinfo {author} {\bibfnamefont {Serge}\ \bibnamefont {Haroche}},\
  }\bibfield  {title} {\enquote {\bibinfo {title} {Real-time quantum feedback
  prepares and stabilizes photon number states},}\ }\href {\doibase
  10.1038/nature10376} {\bibfield  {journal} {\bibinfo  {journal} {Nature}\
  }\textbf {\bibinfo {volume} {477}},\ \bibinfo {pages} {73--77} (\bibinfo
  {year} {2011})}\BibitemShut {NoStop}%
\bibitem [{\citenamefont {Shannon}(1949)}]{shannon_communication_1949}%
  \BibitemOpen
  \bibfield  {author} {\bibinfo {author} {\bibfnamefont {C.~E.}\ \bibnamefont
  {Shannon}},\ }\bibfield  {title} {\enquote {\bibinfo {title} {Communication
  in the {Presence} of {Noise}},}\ }\href {\doibase 10.1109/JRPROC.1949.232969}
  {\bibfield  {journal} {\bibinfo  {journal} {Proceedings of the IRE}\ }\textbf
  {\bibinfo {volume} {37}},\ \bibinfo {pages} {10--21} (\bibinfo {year}
  {1949})}\BibitemShut {NoStop}%
\bibitem [{Note1()}]{Note1}%
  \BibitemOpen
  \bibinfo {note} {Traditional AM radio receiver antennas in cars (meter scale)
  are a good example, which operate with approximately $5\%$ efficiency to
  reach 10~kHz bandwidth with a $\approx $1~MHz carrier.}\BibitemShut {Stop}%
\bibitem [{\citenamefont {Zhu}\ and\ \citenamefont
  {Ziolkowski}(2012)}]{zhu_broad-bandwidth_2012}%
  \BibitemOpen
  \bibfield  {author} {\bibinfo {author} {\bibfnamefont {N.}~\bibnamefont
  {Zhu}}\ and\ \bibinfo {author} {\bibfnamefont {R.~W.}\ \bibnamefont
  {Ziolkowski}},\ }\bibfield  {title} {\enquote {\bibinfo {title}
  {Broad-{Bandwidth}, {Electrically} {Small} {Antenna} {Augmented} {With} an
  {Internal} {Non}-{Foster} {Element}},}\ }\href {\doibase
  10.1109/LAWP.2012.2219572} {\ \textbf {\bibinfo {volume} {11}},\ \bibinfo
  {pages} {1116--1120} (\bibinfo {year} {2012})}\BibitemShut {NoStop}%
\bibitem [{\citenamefont {Bason}\ \emph {et~al.}(2010)\citenamefont {Bason},
  \citenamefont {Tanasittikosol}, \citenamefont {Sargsyan}, \citenamefont
  {Mohapatra}, \citenamefont {Sarkisyan}, \citenamefont {Potvliege},\ and\
  \citenamefont {Adams}}]{bason_enhanced_2010}%
  \BibitemOpen
  \bibfield  {author} {\bibinfo {author} {\bibfnamefont {M.~G.}\ \bibnamefont
  {Bason}}, \bibinfo {author} {\bibfnamefont {M.}~\bibnamefont
  {Tanasittikosol}}, \bibinfo {author} {\bibfnamefont {A.}~\bibnamefont
  {Sargsyan}}, \bibinfo {author} {\bibfnamefont {A.~K.}\ \bibnamefont
  {Mohapatra}}, \bibinfo {author} {\bibfnamefont {D.}~\bibnamefont
  {Sarkisyan}}, \bibinfo {author} {\bibfnamefont {R.~M.}\ \bibnamefont
  {Potvliege}}, \ and\ \bibinfo {author} {\bibfnamefont {C.~S.}\ \bibnamefont
  {Adams}},\ }\bibfield  {title} {\enquote {\bibinfo {title} {Enhanced electric
  field sensitivity of rf-dressed {Rydberg} dark states},}\ }\href {\doibase
  10.1088/1367-2630/12/6/065015} {\bibfield  {journal} {\bibinfo  {journal}
  {New J. Phys.}\ }\textbf {\bibinfo {volume} {12}},\ \bibinfo {pages} {065015}
  (\bibinfo {year} {2010})}\BibitemShut {NoStop}%
\bibitem [{\citenamefont {Degen}\ \emph {et~al.}(2017)\citenamefont {Degen},
  \citenamefont {Reinhard},\ and\ \citenamefont
  {Cappellaro}}]{degen_quantum_2017}%
  \BibitemOpen
  \bibfield  {author} {\bibinfo {author} {\bibfnamefont {C.~L.}\ \bibnamefont
  {Degen}}, \bibinfo {author} {\bibfnamefont {F.}~\bibnamefont {Reinhard}}, \
  and\ \bibinfo {author} {\bibfnamefont {P.}~\bibnamefont {Cappellaro}},\
  }\bibfield  {title} {\enquote {\bibinfo {title} {Quantum sensing},}\ }\href
  {\doibase 10.1103/RevModPhys.89.035002} {\bibfield  {journal} {\bibinfo
  {journal} {Rev. Mod. Phys.}\ }\textbf {\bibinfo {volume} {89}},\ \bibinfo
  {pages} {035002} (\bibinfo {year} {2017})}\BibitemShut {NoStop}%
\bibitem [{\citenamefont {Cox}\ \emph {et~al.}(2016)\citenamefont {Cox},
  \citenamefont {Greve}, \citenamefont {Weiner},\ and\ \citenamefont
  {Thompson}}]{cox_deterministic_2016}%
  \BibitemOpen
  \bibfield  {author} {\bibinfo {author} {\bibfnamefont {Kevin~C.}\
  \bibnamefont {Cox}}, \bibinfo {author} {\bibfnamefont {Graham~P.}\
  \bibnamefont {Greve}}, \bibinfo {author} {\bibfnamefont {Joshua~M.}\
  \bibnamefont {Weiner}}, \ and\ \bibinfo {author} {\bibfnamefont {James~K.}\
  \bibnamefont {Thompson}},\ }\bibfield  {title} {\enquote {\bibinfo {title}
  {Deterministic {Squeezed} {States} with {Collective} {Measurements} and
  {Feedback}},}\ }\href {\doibase 10.1103/PhysRevLett.116.093602} {\bibfield
  {journal} {\bibinfo  {journal} {Phys. Rev. Lett.}\ }\textbf {\bibinfo
  {volume} {116}},\ \bibinfo {pages} {093602} (\bibinfo {year}
  {2016})}\BibitemShut {NoStop}%
\bibitem [{\citenamefont {Han}\ \emph {et~al.}(2018)\citenamefont {Han},
  \citenamefont {Vogt}, \citenamefont {Gross}, \citenamefont {Jaksch},
  \citenamefont {Kiffner},\ and\ \citenamefont {Li}}]{han_coherent_2018}%
  \BibitemOpen
  \bibfield  {author} {\bibinfo {author} {\bibfnamefont {Jingshan}\
  \bibnamefont {Han}}, \bibinfo {author} {\bibfnamefont {Thibault}\
  \bibnamefont {Vogt}}, \bibinfo {author} {\bibfnamefont {Christian}\
  \bibnamefont {Gross}}, \bibinfo {author} {\bibfnamefont {Dieter}\
  \bibnamefont {Jaksch}}, \bibinfo {author} {\bibfnamefont {Martin}\
  \bibnamefont {Kiffner}}, \ and\ \bibinfo {author} {\bibfnamefont {Wenhui}\
  \bibnamefont {Li}},\ }\bibfield  {title} {\enquote {\bibinfo {title}
  {Coherent {Microwave}-to-{Optical} {Conversion} via {Six}-{Wave} {Mixing} in
  {Rydberg} {Atoms}},}\ }\href {\doibase 10.1103/PhysRevLett.120.093201}
  {\bibfield  {journal} {\bibinfo  {journal} {Phys. Rev. Lett.}\ }\textbf
  {\bibinfo {volume} {120}},\ \bibinfo {pages} {093201} (\bibinfo {year}
  {2018})}\BibitemShut {NoStop}%
\bibitem [{\citenamefont {Gard}\ \emph {et~al.}(2017)\citenamefont {Gard},
  \citenamefont {Jacobs}, \citenamefont {McDermott},\ and\ \citenamefont
  {Saffman}}]{gard_microwave--optical_2017}%
  \BibitemOpen
  \bibfield  {author} {\bibinfo {author} {\bibfnamefont {Bryan~T.}\
  \bibnamefont {Gard}}, \bibinfo {author} {\bibfnamefont {Kurt}\ \bibnamefont
  {Jacobs}}, \bibinfo {author} {\bibfnamefont {R.}~\bibnamefont {McDermott}}, \
  and\ \bibinfo {author} {\bibfnamefont {M.}~\bibnamefont {Saffman}},\
  }\bibfield  {title} {\enquote {\bibinfo {title} {Microwave-to-optical
  frequency conversion using a cesium atom coupled to a superconducting
  resonator},}\ }\href {\doibase 10.1103/PhysRevA.96.013833} {\bibfield
  {journal} {\bibinfo  {journal} {Phys. Rev. A}\ }\textbf {\bibinfo {volume}
  {96}},\ \bibinfo {pages} {013833} (\bibinfo {year} {2017})}\BibitemShut
  {NoStop}%
\bibitem [{\citenamefont {Hattermann}\ \emph {et~al.}(2017)\citenamefont
  {Hattermann}, \citenamefont {Bothner}, \citenamefont {Ley}, \citenamefont
  {Ferdinand}, \citenamefont {Wiedmaier}, \citenamefont {S{\'a}rk{\'a}ny},
  \citenamefont {Kleiner}, \citenamefont {Koelle},\ and\ \citenamefont
  {Fort{\'a}gh}}]{hattermann_coupling_2017}%
  \BibitemOpen
  \bibfield  {author} {\bibinfo {author} {\bibfnamefont {H.}~\bibnamefont
  {Hattermann}}, \bibinfo {author} {\bibfnamefont {D.}~\bibnamefont {Bothner}},
  \bibinfo {author} {\bibfnamefont {L.~Y.}\ \bibnamefont {Ley}}, \bibinfo
  {author} {\bibfnamefont {B.}~\bibnamefont {Ferdinand}}, \bibinfo {author}
  {\bibfnamefont {D.}~\bibnamefont {Wiedmaier}}, \bibinfo {author}
  {\bibfnamefont {L.}~\bibnamefont {S{\'a}rk{\'a}ny}}, \bibinfo {author}
  {\bibfnamefont {R.}~\bibnamefont {Kleiner}}, \bibinfo {author} {\bibfnamefont
  {D.}~\bibnamefont {Koelle}}, \ and\ \bibinfo {author} {\bibfnamefont
  {J.}~\bibnamefont {Fort{\'a}gh}},\ }\bibfield  {title} {\enquote {\bibinfo
  {title} {Coupling ultracold atoms to a superconducting coplanar waveguide
  resonator},}\ }\href {\doibase 10.1038/s41467-017-02439-7} {\bibfield
  {journal} {\bibinfo  {journal} {Nature Communications}\ }\textbf {\bibinfo
  {volume} {8}},\ \bibinfo {pages} {2254} (\bibinfo {year} {2017})}\BibitemShut
  {NoStop}%
\bibitem [{\citenamefont {Chen}\ \emph {et~al.}(2013)\citenamefont {Chen},
  \citenamefont {Beck}, \citenamefont {B{\"u}cker}, \citenamefont {Gullans},
  \citenamefont {Lukin}, \citenamefont {Tanji-Suzuki},\ and\ \citenamefont
  {Vuleti{\'c}}}]{chen_all-optical_2013}%
  \BibitemOpen
  \bibfield  {author} {\bibinfo {author} {\bibfnamefont {Wenlan}\ \bibnamefont
  {Chen}}, \bibinfo {author} {\bibfnamefont {Kristin~M.}\ \bibnamefont {Beck}},
  \bibinfo {author} {\bibfnamefont {Robert}\ \bibnamefont {B{\"u}cker}},
  \bibinfo {author} {\bibfnamefont {Michael}\ \bibnamefont {Gullans}}, \bibinfo
  {author} {\bibfnamefont {Mikhail~D.}\ \bibnamefont {Lukin}}, \bibinfo
  {author} {\bibfnamefont {Haruka}\ \bibnamefont {Tanji-Suzuki}}, \ and\
  \bibinfo {author} {\bibfnamefont {Vladan}\ \bibnamefont {Vuleti{\'c}}},\
  }\bibfield  {title} {\enquote {\bibinfo {title} {All-{Optical} {Switch} and
  {Transistor} {Gated} by {One} {Stored} {Photon}},}\ }\href {\doibase
  10.1126/science.1238169} {\bibfield  {journal} {\bibinfo  {journal}
  {Science}\ }\textbf {\bibinfo {volume} {341}},\ \bibinfo {pages} {768--770}
  (\bibinfo {year} {2013})}\BibitemShut {NoStop}%
\bibitem [{\citenamefont {Jau}\ \emph {et~al.}(2016)\citenamefont {Jau},
  \citenamefont {Hankin}, \citenamefont {Keating}, \citenamefont {Deutsch},\
  and\ \citenamefont {Biedermann}}]{jau_entangling_2016}%
  \BibitemOpen
  \bibfield  {author} {\bibinfo {author} {\bibfnamefont {Y.-Y.}\ \bibnamefont
  {Jau}}, \bibinfo {author} {\bibfnamefont {A.~M.}\ \bibnamefont {Hankin}},
  \bibinfo {author} {\bibfnamefont {T.}~\bibnamefont {Keating}}, \bibinfo
  {author} {\bibfnamefont {I.~H.}\ \bibnamefont {Deutsch}}, \ and\ \bibinfo
  {author} {\bibfnamefont {G.~W.}\ \bibnamefont {Biedermann}},\ }\bibfield
  {title} {\enquote {\bibinfo {title} {Entangling atomic spins with a
  {Rydberg}-dressed spin-flip blockade},}\ }\href {\doibase 10.1038/nphys3487}
  {\bibfield  {journal} {\bibinfo  {journal} {Nature Physics}\ }\textbf
  {\bibinfo {volume} {12}},\ \bibinfo {pages} {71--74} (\bibinfo {year}
  {2016})}\BibitemShut {NoStop}%
\bibitem [{\citenamefont {Li}\ and\ \citenamefont
  {Kuzmich}(2016)}]{li_quantum_2016}%
  \BibitemOpen
  \bibfield  {author} {\bibinfo {author} {\bibfnamefont {Lin}\ \bibnamefont
  {Li}}\ and\ \bibinfo {author} {\bibfnamefont {A.}~\bibnamefont {Kuzmich}},\
  }\bibfield  {title} {\enquote {\bibinfo {title} {Quantum memory with strong
  and controllable {Rydberg}-level interactions},}\ }\href {\doibase
  10.1038/ncomms13618} {\bibfield  {journal} {\bibinfo  {journal} {Nature
  Communications}\ }\textbf {\bibinfo {volume} {7}},\ \bibinfo {pages} {13618}
  (\bibinfo {year} {2016})}\BibitemShut {NoStop}%
\end{thebibliography}%

\end{document}